\documentclass[]{elsart}
\usepackage{graphicx, amssymb, amsfonts,amsmath}
\usepackage{cite}
\usepackage{subfigure}
\usepackage{hyperref}
\usepackage[bf,footnotesize]{caption}
\usepackage{hyperref}
\usepackage{appendix}
\usepackage{cite}

\graphicspath{{plots/}}

\def \3{\ss }

\def\fm{{\rm fm}}

\newcommand{\Oasq}{\mathcal{O}(a^2)}
\newcommand{\mps}{m_\pi}
\newcommand{\fps}{f_\pi}
\newcommand{\mpstm}{m_\pi^{\rm tm}}
\newcommand{\fpstm}{f_\pi^{\rm tm}}
\newcommand{\mpsov}{m_\pi^{\rm ov}}
\newcommand{\fpsov}{f_\pi^{\rm ov}}
\newcommand{\mov}{m_{\rm ov}}
\newcommand{\MSb}{\overline{\mathrm{MS}}}
\newcommand{\mev}{\mathrm{MeV}}
\newcommand{\gev}{\mathrm{GeV}}

\newcommand{\nft}{N_{\rm f}=2}

\journal{}

\begin{document}

\begin{frontmatter}
  
  \vspace*{-1.0truecm} \title{
    Continuum Limit of Overlap Valence Quarks on a Twisted Mass Sea
  }
  \vspace*{-0.5truecm}

  \author[a]{Krzysztof Cichy\corauthref{cor}},
  \ead{kcichy@amu.edu.pl}
  \author[b,c]{Gregorio Herdoiza},
  \ead{gregorio.herdoiza@uam.es}
  \author[b]{Karl Jansen}
  \ead{Karl.Jansen@desy.de}
  
  \corauth[cor]{Corresponding author.}
  
  \address[a]{Adam Mickiewicz University, Faculty of Physics\\
    Umultowska 85, 61-614 Poznan, Poland, tel. +48 618295076, fax. +48 618295070}

  \address[b]{NIC, DESY, Platanenallee 6, D-15738 Zeuthen,
    Germany}
  \address[c]{Departamento de F\'isica Te\'orica, Universidad
    Aut\'onoma de Madrid\\
    Cantoblanco E-28049 Madrid, Spain}

  \begin{abstract}

    \noindent We study a lattice QCD mixed action with overlap valence
    quarks on two flavours of Wilson maximally twisted mass sea
    quarks.  Employing three different matching conditions to relate
    both actions to each other, we investigate the continuum limit by
    using three values of the lattice spacing ranging from $a \approx
    0.05\,\fm$ to $0.08\,\fm$.  A particular emphasis is put on the
    effect on physical observables of the topological zero modes
    appearing in the valence overlap operator. We estimate the region
    of parameter space where the contribution from these zero modes is
    sufficiently small such that their effects can be safely
    controlled and a restoration of unitarity of the mixed action in
    the continuum limit is reached.
    
  \end{abstract}

  \begin{keyword}
    Lattice gauge theory, mixed action, chiral fermions.
    \PACS 11.15.Ha, 12.38.Gc\\
    Preprint-No: DESY 10-240, FTUAM-10-35, SFB/CPP-10-130  \\
  \end{keyword}

\end{frontmatter}

\section{Introduction}

The discretization of the QCD action on a lattice provides a
framework for a 
quantitative description of the non-perturbative phenomena inherent to
hadronic physics at low energies. 
The way of discretizing the continuum QCD action is, however, by no means unique
and leaves a large number of choices. 
Although all lattice actions used in practise respect the principle of 
local gauge invariance, they often break a number of continuum 
symmetries depending on the particular choice of the lattice 
action used. A most notable example is the explicit breaking of
chiral symmetry on the lattice in order to avoid unwanted
additional fermion excitations, the so-called doublers 
\cite{Nielsen:1980rz,Nielsen:1981xu,Friedan:1982nk}. 
There exist, however,  
lattice fermion actions which are constructed from Dirac operators
that satisfy the Ginsparg-Wilson relation \cite{Ginsparg:1981bj} and lead to an 
exactly preserved but lattice modified chiral symmetry \cite{Luscher:1998pqa}. 
The big disadvantage of these Ginsparg-Wilson type fermions is 
that they are 
computationally very expensive.

A possible way to restore a desired symmetry --at least partially,
since it will not by respected by the sea quark action -- is the
approach of so-called mixed actions, corresponding to the use of
different discretizations of the Dirac operator in the valence and sea
sectors.  Of course, using such a mixed action it needs to be
guaranteed that a well defined continuum limit, which corresponds to
the universality class of continuum QCD, is reached eventually. This
can be achieved by imposing an appropriate matching condition relating
both actions to each other and which fixes the physical situation at
each value of a non-zero lattice spacing.  Hence, a well defined
continuum limit can be reached.

Although in this way mixed actions offer an advantage since the
valence quarks can have improved symmetry properties with respect to
the sea quark action, they also provide challenges.  Even when
imposing a matching condition, the sea quark and valence quark
theories are not fully matched and differ by discretization effects,
which lead to unitarity violations in physical quantities.  It is
therefore an important question to understand the characteristics of a
mixed action approach and to learn how and whether indeed the correct
continuum limit is reached. In particular, it is an interesting and
important question, whether and how the inherent mismatch of the
eigenvalue spectra affect physical observables and influence possibly
the continuum limit.

It is the goal of this work  
to address the above  
questions by studying a particular mixed action setup, 
employing maximally twisted mass (MTM) fermions 
\cite{Frezzotti:2000nk,Frezzotti:2003ni} in the sea 
and chirally invariant overlap fermions \cite{Neuberger:1997fp,Neuberger:1998wv} in the valence 
sector. As anticipated above, the 
mismatch of the spectra of the lattice Dirac operators employed 
in the sea and the valence actions will be a particular concern. 
While the overlap Dirac operator exhibits exact chiral zero modes, 
for any non-zero value of the lattice spacing, 
the twisted mass Dirac operator does not, 
at least at lattice 
spacings used in this work. 
As discussed already 
in ref.~\cite{Blum:2000kn}, the quantitative effects of these unmatched zero modes 
on physical observables 
will depend on the quark mass and the physical volume 
used in the numerical simulation. 

We will investigate the mesonic sector 
of the theory and study 
the continuum limit of light pseudoscalar meson observables 
by using three values of the lattice spacing ranging from $a
\approx 0.05\,\fm$ to $0.08\,\fm$ at a fixed lattice size of 
$L \approx 1.3\,\fm$, and quark masses
$m_{q}^{\MSb}(\mu=2\,\gev)\approx 20\,\mev$ and 
$m_{q}^{\MSb}(\mu=2\,\gev)\approx 40\,\mev$. 
These values of the quark masses correspond to infinite volume 
pseudoscalar masses of approximately $300\, \mev$ and $450\, \mev$, respectively. 
We also study finite size effects by considering three
values of the lattice size between $L\approx1.3\,\fm$ and $2.0\,\fm$. The
choice 
of this parameter 
region (quark mass, lattice size
and lattice spacing) allows us to perform then a systematic study of our 
mixed action setup. 

Mixed actions are currently being used by various groups. In the case
of valence overlap quarks, studies using domain wall sea
fermions~\cite{Allton:2006nu,Li:2010pw} or Wilson clover sea
quarks~\cite{Durr:2007ef,Bernardoni:2010nf} have been carried out.
Domain wall valence fermions on staggered sea quarks have also been
extensively used~\cite{Renner:2004ck,Beane:2006kx,Aubin:2008wk}. Mixed
actions in which variants of the sea action are used in the valence
sector, such as Osterwalder-Seiler fermions on a twisted mass sea,
have also been recently
studied~\cite{Constantinou:2010qv,Urbach:lat10,Herdoiza:lat10}. 
The extension of chiral perturbation theory to the context of mixed
actions~\cite{Bar:2002nr,Golterman:2005xa,Chen:2007ug} provides a
useful analytic control of the light quark mass dependence of several
physical quantities in the regime of parameters (quark mass, lattice
size and lattice spacing) where the effective theory can be applied.

As said above, 
in this paper, we perform a systematic study of a mixed action with
overlap valence fermions on a Wilson twisted mass sea, by focusing on
light pseudoscalar meson observables. In Section 2 we introduce the
mixed action setup by shortly reviewing the overlap and twisted mass
formalisms and introducing the quark mass matching procedure, which is
illustrated in the free theory.  
Section 3 describes the simulation setup
and Section 4 presents the results of the continuum limit study of the
mixed action for our light quark mass ensembles. 
Section 5 provides an interpretation and discussion of the results.
In Section 6 we discuss the
regime of heavier quark masses and finite size effects.
Section 7 concludes and provides an estimate of the
regime of parameters where the contribution in the valence sector of
the topological zero modes on physical observables 
is small enough such that a proper control
of the restoration of unitarity of the mixed action becomes possible.

\section{Mixed Action: Fermionic Actions, Matching and Illustration in the Free Theory}

\subsection{Overlap Fermions}
Overlap fermions were introduced in 1997 by Neuberger
\cite{Neuberger:1997fp,Neuberger:1998wv}, who found a particularly simple form
of a lattice Dirac operator that obeys the Ginsparg-Wilson relation
\cite{Ginsparg:1981bj}. This implies that overlap (ov) fermions preserve a
lattice version of chiral symmetry at any value of the lattice spacing
\cite{Luscher:1998pqa}.  The massless overlap Dirac operator is given
by\,\footnote{For a review about overlap fermions, see
  e.g. \cite{Niedermayer:1998bi}.}:
\begin{equation}
  \label{massless_overlap}
  \hat D_{\rm ov}(0)=\frac{1}{a}\Big(1-A(A^\dagger A)^{-1/2}\Big).
\end{equation}
The kernel operator $A$ reads:
\begin{equation}
  \label{overlap-A}
  A=1+s-a\hat D_{\rm Wilson}(0),
\end{equation}
where $s$ is a parameter which satisfies $|s|<1$. The Wilson-Dirac operator is defined by:
\begin{equation}
  \label{eq:DW}
  \hat D_{\rm Wilson}(m_0)=\frac{1}{2}\left(\gamma_\mu(\nabla_\mu^*+\nabla_\mu)-a\nabla_\mu^*\nabla_\mu\right)+m_0,
\end{equation}
where $m_0$ is the bare Wilson quark mass and $\nabla_\mu$, $\nabla^*_\mu$ are the forward and backward covariant derivatives, respectively.
The massive overlap Dirac operator is given by:
\begin{equation}
  \label{overlap-massive}
  \hat D_{\rm ov}(m_{\rm ov}) = \left(1-\frac{am_{\rm ov}}{2}\right)\hat D_{\rm ov}(0)+m_{\rm ov}, \nonumber
\end{equation} 
where $m_{\rm ov}$ is the bare overlap quark mass.
Note that despite the appearance of the square root in eq.~(\ref{massless_overlap}), the overlap 
operator is local in the sense of exponentially decaying coupling strengths  
as a function of the distance between lattice points \cite{Hernandez:1998et}.

Operators preserving the Ginsparg-Wilson relation develop topological
zero modes and fulfil the index theorem at finite lattice
spacing~\cite{Hasenfratz:1998ri}. Moreover, chiral symmetry allows to
improve the theory at $\mathcal{O}(a)$. Indeed, observables computed
with overlap fermions are not affected by $\mathcal{O}(a)$ lattice
artefacts, provided that $\mathcal{O}(a)$-improved interpolating
operators are used (see e.g.~\cite{Bietenholz:2004wv}).

The overlap quark mass $\mov$ renormalizes multiplicatively with a
renormalisation factor $Z_m=1/Z_{\rm P}$, where P stands for the
pseudoscalar density. The decay constant $\fps$ of a mass degenerate
pseudoscalar meson, carrying a mass $\mps$, can be determined by means
of the two-point pseudoscalar correlation function $C_{\rm PP}(t)$ via
the following expression:
\begin{equation}
  \label{fpsov}
  \fpsov = \frac{2\mov}{ ( \mpsov )^2} \, |\langle 0| P |\pi \rangle_{\rm ov} | \,.  \nonumber
\end{equation}

It is known that in 
practise, overlap fermions are demanding from the computational point
of view. With respect to Wilson type fermions the overall cost can
grow by one to two orders of magnitude\,\footnote{We refer
  to~\cite{Chiarappa:2006hz} for a detailed comparison to Wilson twisted mass
  fermions.}. Moreover, the discontinuities of the overlap operator
while changing topological sectors pose a problem to the standard
Molecular Dynamics algorithm used in dynamical quark
simulations. Modifications of the Molecular Dynamics evolution to
tackle this
problem~\cite{Fodor:2003bh,Cundy:2005mr,Schaefer:2006bk,Cundy:2008zc} imply
a further significant increase in the computational cost. A way to
avoid the discontinuity problem is to modify the action in such a way
that the topological charge remains
fixed~\cite{Fukaya:2006vs,Aoki:2008tq}. However, this procedure introduces
additional $\mathcal{O}(1/V)$ finite volume effects which need to be
taken into account in the determination of physical observables.

\subsection{Wilson Twisted Mass Fermions}

Twisted mass (tm) fermions~\cite{Frezzotti:2000nk} are defined by adding a chirally rotated mass
term to the Wilson-Dirac operator in eq.~(\ref{eq:DW}) as follows:
\begin{equation}
  \label{eq:Dtm}
  \hat D_{\rm tm} = \hat D_{\rm Wilson}(m_0)+i\mu_q\gamma_5\tau_3\,,
\end{equation} 
where $\mu_q$ is the twisted mass parameter and $\tau_3$ is the third
Pauli matrix acting in flavour space.

Wilson twisted mass fermions were introduced to address the problem of
unphysically small eigenvalues of the Wilson-Dirac
operator~\cite{Frezzotti:2000nk}. Furthermore, by tuning the mass $m_0$ to
its critical value $m_{\rm crit}$, a situation denoted by maximal
twist, physical observables are automatically $\mathcal{O}(a)$
improved \cite{Frezzotti:2003ni} independently of the operator which is
considered, implying that no additional, operator specific improvement
coefficients need to be computed. Detailed studies of the
continuum-limit scaling in the quenched
approximation~\cite{Jansen:2005kk} and with two dynamical
quarks~\cite{Baron:2009wt} have demonstrated that, after an
appropriate tuning procedure to maximal twist, lattice artefacts
indeed follow the expected $\Oasq$ scaling behaviour.

A comprehensive simulation programme has been undertaken by the
European Twisted Mass (ETM) collaboration, including dynamical
simulations with $N_{\rm f}=2$ \cite{Boucaud:2007uk,Boucaud:2008xu,Baron:2009wt}
and $N_{\rm f}=2+1+1$~\cite{Baron:2010bv} quark flavours.

At maximal twist, the quark mass is given by the twisted mass $\mu_q$
and, as in the case of overlap fermions, it renormalizes
multiplicatively with a renormalisation factor $Z_\mu=1/Z_{\rm P}$.
In contrast to standard Wilson fermions, an exact lattice Ward
identity for maximally twisted mass fermions allows the extraction of
the decay constant of the charged pseudoscalar meson from the relation
\begin{equation}
  \label{fpstm}
  \fpstm = \frac{2\mu_q}{(\mpstm)^2} \,  |\langle 0| P|\pi^\pm \rangle_{\rm tm} |\,, \nonumber
\end{equation}
which is free from the presence of renormalisation factors.

The twisted mass term in eq.(\ref{eq:Dtm}) explicitly breaks parity
and isospin symmetry, which are however restored in the continuum
limit with a rate of $\Oasq$ as shown in~\cite{Frezzotti:2003ni} and
numerically confirmed in~\cite{Jansen:2005cg,Baron:2009wt}. The
effect of isospin breaking has been observed to affect in a
substantial way only the neutral pion mass. Indeed, while the
discretization effects in the charged pion are observed to be small,
significant $\mathcal{O}(a^2)$ corrections appear when studying the
scaling to the continuum limit of the neutral
pion~\cite{Baron:2009wt}. Similar effects have not been observed in
other quantities that are in principle sensitive to isospin breaking
but not trivially related to the neutral pion mass. These observations
are supported by theoretical considerations detailed
in~\cite{Frezzotti:2007qv,Dimopoulos:2009qv}.

\subsection{Matching Procedure}
\label{sec:mat}

In order to perform the continuum limit extrapolation, the physical
situation (e.g. the quark mass and the lattice size) has to be kept
fixed when changing the lattice spacing. 
In practise, this requires a matching procedure at each
value of the lattice spacing. One possibility for this matching is to
directly match the renormalized quark masses. Another option is to
match a hadronic quantity that can be accurately determined with both
regularisation and that depends in a significant way on the quark
mass. In the light quark sector, the mass of the pseudo Nambu-Goldstone
boson is a natural hadronic observable to perform this matching. In
this work we follow this procedure, by defining the matching point in
the following way:
\begin{equation}
  \label{eq:match}
  \mpsov|_{\mov\equiv\mov^{\rm match}}=\mpstm|_{\mu_q=\mu_q^{\rm sea}}\,,
\end{equation}
where the l.h.s is the mass of a pseudoscalar meson consisting of 
mass degenerate overlap valence quarks of mass $\mov=\mov^{\rm match}$ (the so
called matching quark mass) in a background of $\nft$ twisted mass sea
quarks with mass $\mu_q=\mu_q^{\rm sea}$. The mass $\mpstm$ in the
r.h.s corresponds to the pseudoscalar meson made of twisted mass
valence and sea quarks of mass $\mu_q=\mu_q^{\rm
  sea}$. 

To illustrate the matching procedure in the valence sector, we perform
a continuum limit scaling analysis of overlap and twisted mass
fermions in the free theory.

\subsubsection*{Illustration in the Free Theory}

We want to investigate the effects of the matching procedure of the
quark mass in the free theory by using as a matching condition the
pseudoscalar meson masses of maximally twisted mass fermions and of
overlap fermions. The twisted mass is kept fixed to $N\mu_q=0.5$,
where $N$ is the number of lattice sites in the spatial directions.

\begin{figure}[t]
  \centering
  \subfigure[\label{fig-tl-pionmass-matching}]{
    \includegraphics[width=0.33\linewidth,angle=270]{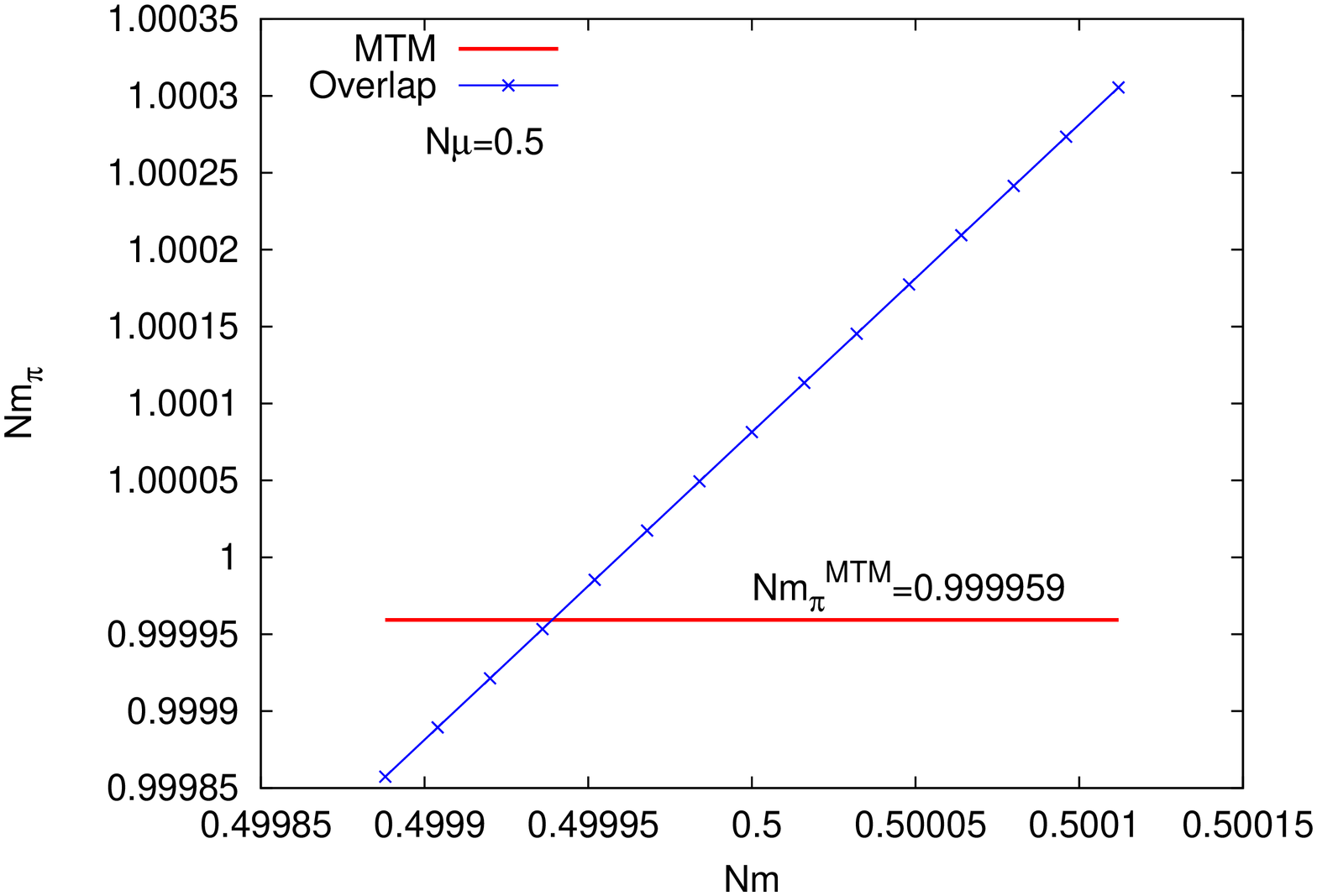}
  }
  \subfigure[\label{fig-tl-piondecay-matching}]{
    \includegraphics[width=0.33\linewidth,angle=270]{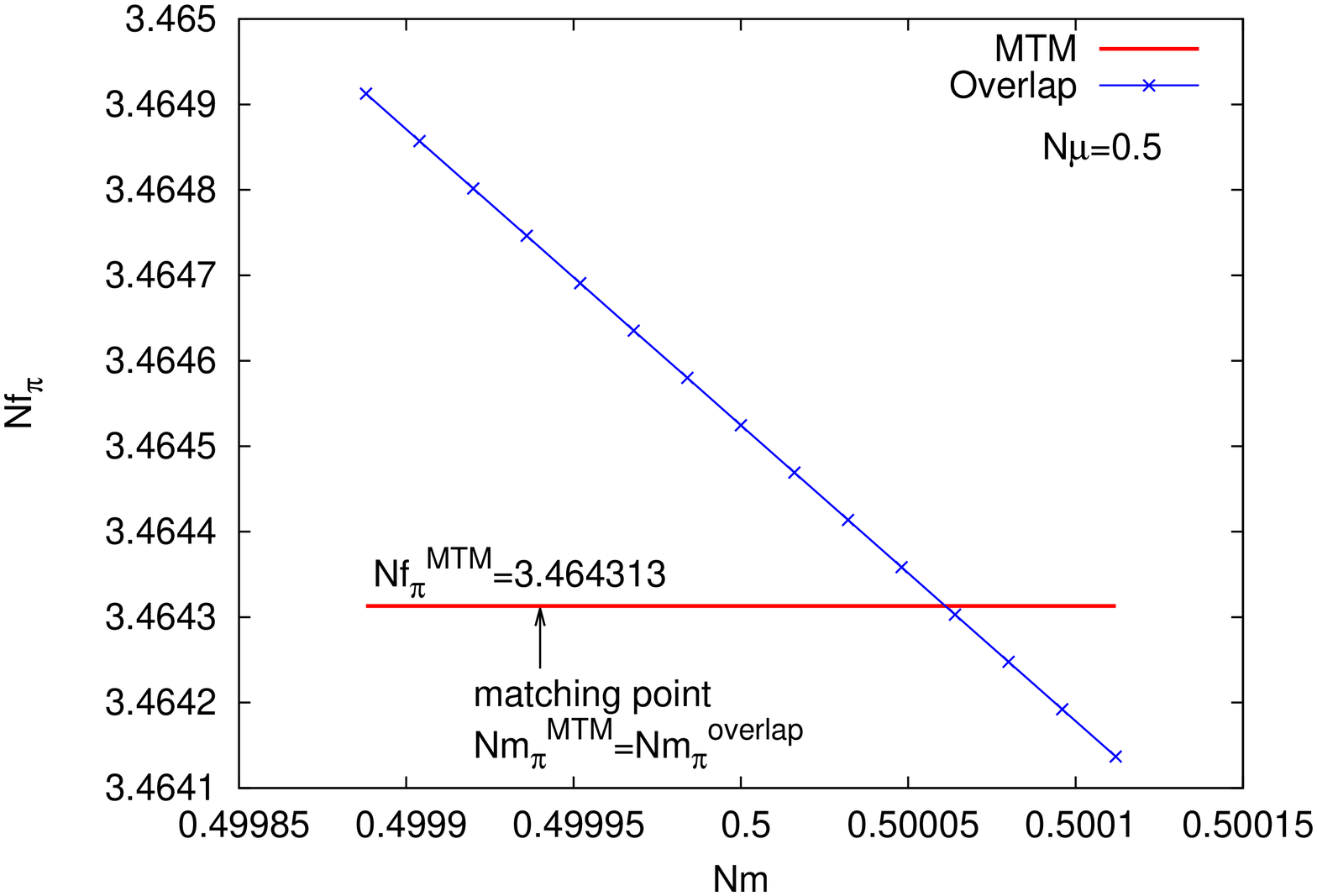}
  }
  \caption{Tree-level examples~: (a) the matching of twisted mass and
    overlap quark masses and (b) the mismatch between the twisted mass
    and overlap pseudoscalar decay constants at the matching point
    $N\mpstm=N\mpsov$.}
\end{figure}

We illustrate the matching procedure through an example with $N=16$.
In this case, the twisted mass pseudoscalar meson mass is $N\mpstm=0.999959$.
Since the magnitude of $\mathcal{O}(a^2)$
effects is in general different for different fermion discretizations,
if we impose the condition of equal pseudoscalar meson masses $N\mpstm=N\mpsov$,
the bare quark masses will differ $N \mov \neq N\mu_q$. This is shown in Fig.
\ref{fig-tl-pionmass-matching}, where the dependence of the overlap
pseudoscalar meson mass on the quark mass $N\mov$ is depicted. The
value $N\mov^{\rm match}\approx0.49994$ leads to the same pseudoscalar
meson mass as the value $N\mu_q=0.5$ in the twisted mass case.

Since the size of discretization effects depend on the particular
observables under consideration, the matching of the pseudoscalar
meson mass does not imply that other observables will also be matched. This
is illustrated in Fig. \ref{fig-tl-piondecay-matching} which shows the
difference of the pseudoscalar decay constants at the matching point,
$N\mpstm=N\mpsov$.

\begin{figure}[t]
  \begin{center}
    \includegraphics[width=0.45\textwidth,angle=270]{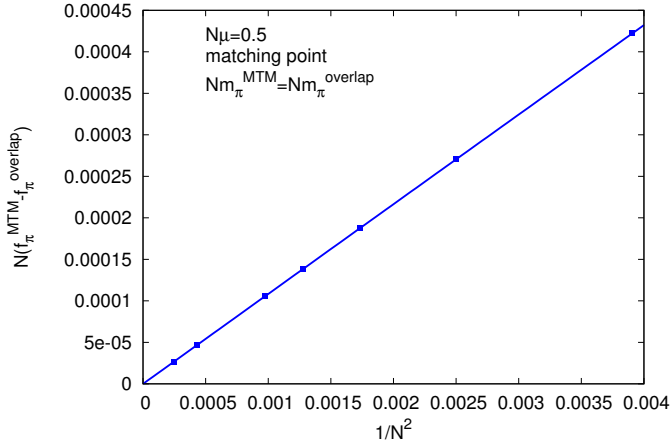}
    \caption{The difference, in the free theory, between the twisted
      mass and overlap pseudoscalar meson decay constants at the
      matching point $N\mpstm=N\mpsov$, as a
      function of $1/N^2$. The limit $1/N^2 \to 0$ corresponds to the
      continuum limit.}
    \label{fig-tl-piondecay-matching-scaling}
  \end{center}
\end{figure}

The difference $N\fpstm-N\fpsov$ vanishes in
the continuum limit at a rate of $\mathcal{O}(a^2)$. We consider
different lattice sizes, by changing $N$ in the range $N=4$ to
$N=64$. By fixing the physical situation to $N\mu_q=0.5$, the change
in $N$ introduces the scaling towards the continuum limit, which, in
the free theory, corresponds to the limit
$N\rightarrow\infty$\,.\footnote{More details about tree-level scaling
  tests of various fermion discretizations and their expressions for
  the quark propagators and correlation functions can be found in
  \cite{Cichy:2008gk,Cichy:2008nt}.} For each lattice size $N$, we determine the
matching overlap quark mass $N\mov^{\rm match}$, such that the overlap
and twisted mass pseudoscalar meson masses are equal. In
Fig.~\ref{fig-tl-piondecay-matching-scaling} we illustrate the
dependence of the difference $Nf_{\rm PS}^{\rm tm}-Nf_{\rm PS}^{\rm
  ov}$ on $1/N^2$. This difference indeed vanishes in the limit
$N\rightarrow\infty$ and the leading discretization effects are of
$\mathcal{O}(1/N^2)$. A similar behaviour is observed for other
observables such as the pseudoscalar correlation function at a fixed
physical distance.

\section{Simulation Setup}

In the following, we will investigate the just discussed 
example of free fermions in the case of including the interaction to the gluon fields.
In table~\ref{tab:setup} we provide the list of dynamical simulations
considered in this work. The ensembles were generated by the ETM
collaboration~\cite{Baron:2009wt} using the tree-level Symanzik
improved gauge action and $\nft$ flavours of maximally twisted mass
fermions.

The heavier ensembles labelled in table~\ref{tab:setup} with a
subscript $h$, i.e. $B_h$, $C_h$ and $D_h$, correspond to three values
of the lattice spacing with a physical situation fixed to a lattice
size $L/r_0 \approx 3$ and a charged pseudoscalar meson mass $\mpstm
r_0~\approx~1.0$, where $r_0$ is the Sommer
scale~\cite{Sommer:1993ce}. This corresponds to a lattice size $L
\approx 1.3\,\fm$ and a non-perturbatively
renormalized~\cite{Constantinou:2010gr} quark mass
$\mu_{q}^{\MSb}(\mu=2\,\gev)\approx 40\,\mev$. In infinite volume this
corresponds to a pseudoscalar meson mass $\mps \approx 450\,\mev$.

The lighter ensembles are labelled with a subscript $\ell$. In this
case the pseudoscalar meson mass is fixed to $\mps r_0 \approx 0.8$.
This corresponds to a quark mass $\mu_{q}^{\MSb}(\mu=2\,\gev)\approx
20\,\mev$, which in infinite volume gives a pseudoscalar meson mass of
around $300\,\mev$. The subscripts $V_{1,2,3}$ refer to ensembles with
different spatial lattice extents.

\begin{table}[t!]
  \centering
  \begin{tabular*}{1.0\textwidth}{@{\extracolsep{\fill}}lcccccc}
    \hline\hline
    Ensemble & $\beta$ & $(L/a)^3\times T/a$  & $a\mu_q$ & $\kappa_\mathrm{crit}$ &
    $\mpstm r_0$ & $L/r_0$\\
    \hline\hline
    $B_h$      &  $3.90$ & $16^3\times 32$ & $0.0074$ & $0.160856$ & $1.03$  & $3.0$ \\
    $B_{\ell,\,V_1}$  &  & $16^3\times 32$ & $0.0040$ &            & $0.84$  &       \\
    $B_{\ell,\,V_2}$    && $20^3\times 40$    &          &            & $0.73$  & $3.8$ \\
    $B_{\ell,\,V_3}$    && $24^3\times 48$    &          &            & $0.72$  & $4.6$ \\
    \hline
    $C_h$      & $4.05$  & $20^3\times 48$ & $0.0060$ & $0.157010$ & $1.00$ & $3.0$ \\
    $C_{\ell}$          && $20^3\times 40$ & $0.0030$ &            & $0.83$ &       \\
    \hline
    $D_h$      & $4.20$  & $24^3\times 48$ & $0.0050$ & $0.154073$ & $1.04$ & $2.9$ \\
    $D_{\ell}$          && $24^3\times 48$ & $0.0020$ &            & $0.82$ &       \\
    \hline\hline\\
  \end{tabular*}
  \caption{Summary of $\nft$ Wilson twisted mass ensembles considered
    in this work. We give the ensemble label, the values of the
    inverse coupling $\beta$, the lattice volume $L^3\times T$, the
    twisted mass parameter $a\mu_q$, the critical hopping parameter
    $\kappa_\mathrm{crit} = 1/(8+ 2am_{\rm crit})$, the approximate
    values of the pseudoscalar meson mass and of the lattice size in
    units of the Sommer scale $r_0$.  The chirally extrapolated values
    of the Sommer scale are $r_0/a=5.25(2)$ at $\beta=3.90$,
    $r_0/a=6.61(2)$ at $\beta=4.05$ and $r_0/a=8.33(5)$ at
    $\beta=4.20$, corresponding to lattice spacings $a\approx0.079\,
    \mathrm{fm}$, $a\approx0.063\, \mathrm{fm}$ and $a\approx0.051\,
    \mathrm{fm}$, respectively~\cite{Baron:2009wt}.  The subscripts
    $h$ and $\ell$ in the ensemble name refer to the heavy and light
    quark masses, respectively, which are kept fixed when performing
    the continuum limit extrapolation. The subscripts $V_{1,2,3}$ label
    ensembles with different lattice
    sizes.}
  \label{tab:setup}
\end{table}

In the valence sector we use overlap fermions\,\footnote{More details about our overlap fermions setup are given in \cite{Cichy:2009dy}.}. The gauge links
entering in the covariant derivative of the kernel operator $A$ in
eq.~(\ref{overlap-A}) are HYP smeared to reduce the condition number
of $A^\dagger A$. The mass parameter $s$ is set to $s=0$ since this is observed
to provide the best locality properties for the overlap
operator~\footnote{Preliminary results were presented in
  \cite{Cichy:lat10} and we refer to a forthcoming
  publication~\cite{Drach:2010} for more details about this aspect.}.

For each dynamical quark ensemble in table~\ref{tab:setup}, we produce
all-to-all overlap quark propagators in a wide range of quark masses
$m_{\rm ov}$, from the unitary (sea) light quark mass up to the
physical strange quark mass. This provides a good handle on the
matching procedure, to be discussed in the following section.
Note that the lattice size $L/r_0 \approx 3$ is outside the regime of
applicability of chiral perturbation theory (or its generalisation to
mixed actions). In this work, we instead rely on a direct study of the
continuum limit of the mixed action in order to characterise the size
of unitarity violations.

\section{Continuum Limit Scaling Tests and Different Matching Conditions}

In this section, we will discuss a number of
continuum limit scaling tests for the pseudoscalar decay constant.
We will employ three different matching conditions 
to relate the overlap valence quarks to the maximally twisted 
mass sea quarks and the pseudoscalar decay constant will be extracted
from two different correlation functions which differ by their sensitivity 
to chiral zero modes. 
We remark that we consider the matching condition employed {\em to define} 
the mixed action theory used. This is similar to improvement conditions where the 
resulting improvement coefficients define the improved actions and operators. 
In particular, there is then no uncertainty associated with a given matching
condition. However, it is of course still important to investigate different matching 
conditions and study their effects, as we will do in this work. 

In the following we will use the shorthand notation ``scaling test" for the 
continuum limit scaling test in the lattice spacing. 
As in the example of free fermions discussed above, we start with a {\em naive scaling test}. 
Demonstrating that such naive procedure leads to a problem, when 
the continuum limit is taken,  
we will introduce an {\em improved scaling test}. 
By also discussing an {\em alternative scaling test}, 
we will then demonstrate that physical results and the approach 
to the continuum limit can strongly depend on the choice 
of the setup for a scaling test.
We provide an explanation for this finding by the effects of 
exact chiral zero modes of the overlap Dirac operator taken in the valence
sector.

\subsection{Naive Scaling Test}
\label{sec:naive}

The {\em naive scaling test} follows very closely the example 
of free fermions discussed above. 
As in the free case, we match the pseudoscalar meson masses computed from the correlator 
constructed from the pseudoscalar interpolating field (PP correlator, $C_{\rm PP}(t)$), 
which appears to be the most natural choice.
Of course, in practical simulations, the pseudoscalar masses cannot
not be matched exactly. We therefore use that value of the overlap pseudoscalar 
mass which is closest to the targeted twisted mass value.
The corresponding overlap bare quark mass then defines our matching quark 
mass.
Using $C_{\rm PP}(t)$ for the matching condition we will refer to as {\em naive matching condition}.
To proceed, we evaluate the pseudoscalar correlation function 
$C_{\rm PP}(t)$
for light quark masses using 
ensembles $B_{\ell,V_1}$,
$C_\ell$ and $D_\ell$.

\begin{table}[t!]
  \centering
  \begin{tabular*}{0.8\textwidth}{@{\extracolsep{\fill}}lcccc}
    \hline\hline
    Ensemble $\downarrow$ & $a\mps$ & $a\mov^{\rm match}$ & $a\mov^{\rm match}$ & $a\mov^{\rm match}$\\
    matching $\rightarrow$ & & naive & improved & alternative\\
    \hline\hline
    $B_h$      & 0.1961(17) & 0.015 & 0.016 & 0.017\\ 
    $B_{\ell,\,V_1}$  & 0.1592(24) & 0.007 & 0.011 & 0.009\\ 
    $B_{\ell,\,V_2}$    & 0.1389(14)& 0.007 & 0.009 & --- \\
    $B_{\ell,\,V_3}$    & 0.1359(7) & 0.008 & 0.008 & --- \\
    \hline
    $C_h$      & 0.1520(15) & 0.011 & 0.012 & 0.013 \\ 
    $C_{\ell}$          & 0.1209(40) & 0.005 & 0.006 & 0.006 \\ 
    \hline
    $D_h$      & 0.1252(13) & 0.009  & 0.010 & 0.010 \\ 
    $D_{\ell}$          & 0.0980(19) & 0.002 & 0.004 & 0.004 \\ 
    \hline\hline\\
  \end{tabular*}
  \caption{The pseudoscalar masses and overlap matching quark masses for different ensembles. Results of employing different matching conditions are shown.}
  \label{tab:mat}
\end{table}

\begin{figure}[t]
  \centering
  \subfigure[\label{fig:matL}]{
    \includegraphics[width=0.33\linewidth,angle=270]{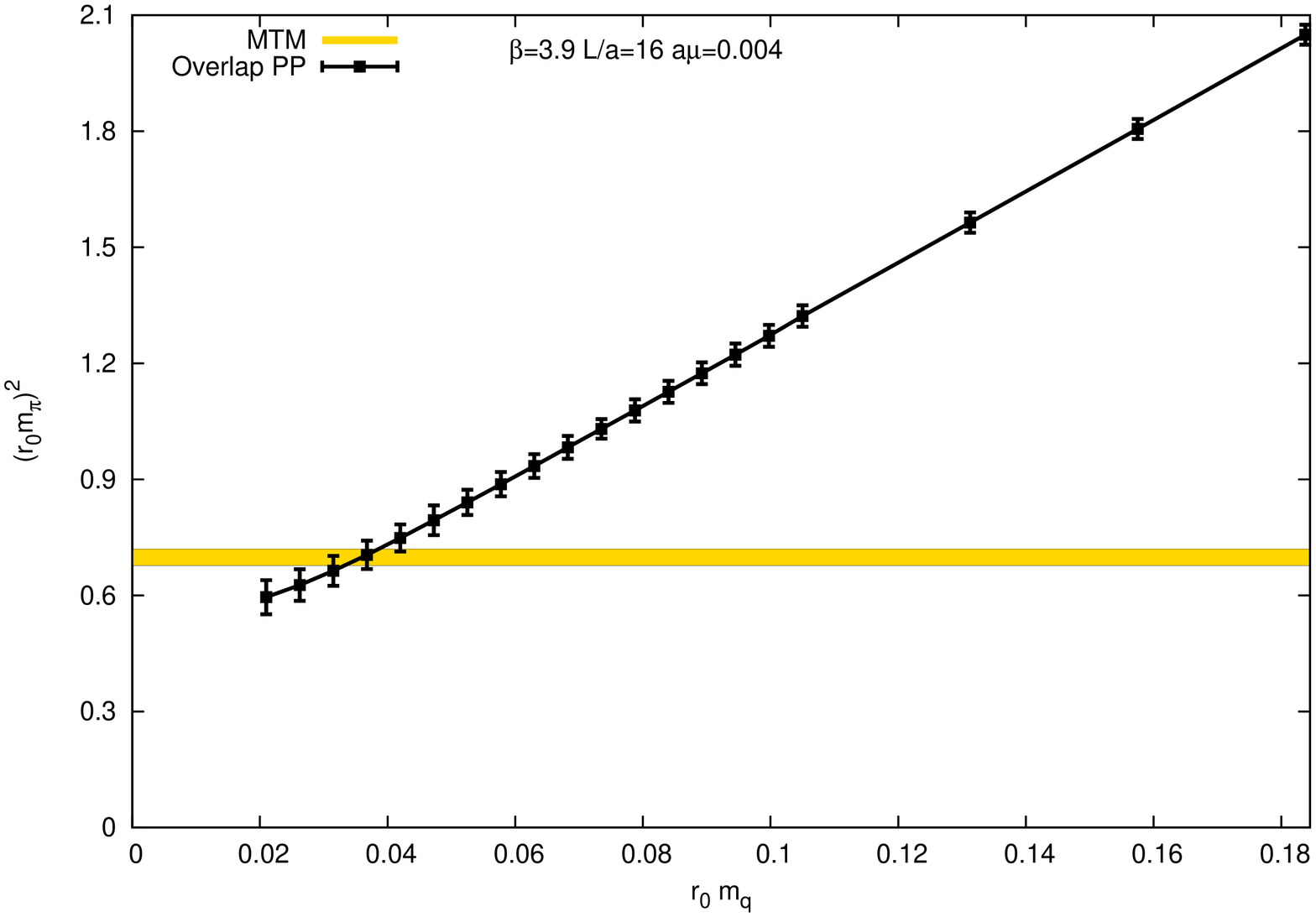}
  }
  \subfigure[\label{fig:fpsLPS}]{
    \includegraphics[width=0.33\linewidth,angle=270]{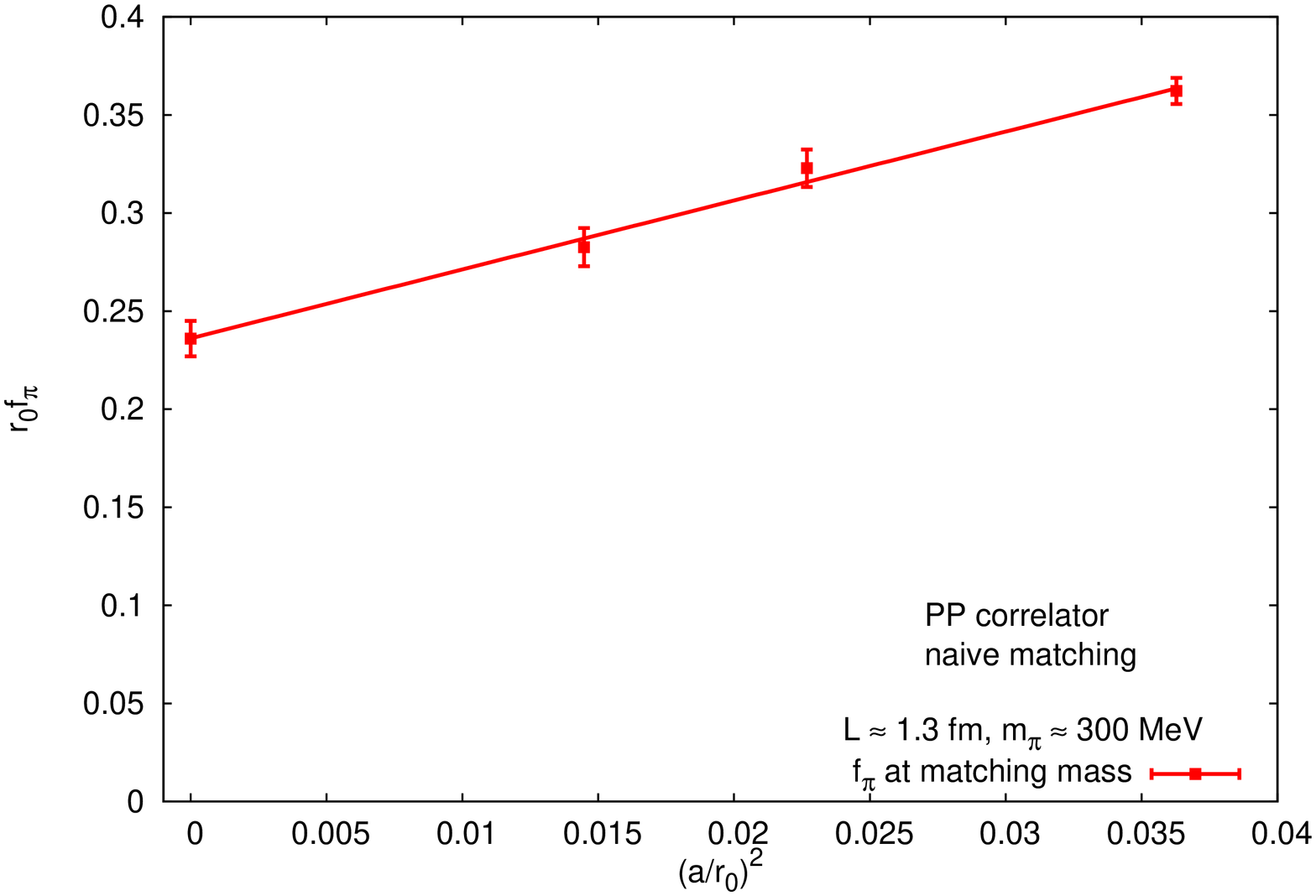}
  }
  \caption{(a) Naive matching condition for the case of ensemble $B_{\ell,V_1}$. (b) Continuum limit of $\fps$ in the case of the naive matching condition.  }
\end{figure}

The matching procedure is illustrated for the case $B_{\ell,V_1}$ in Fig.~\ref{fig:matL}. 
The values of the matching quark and pseudoscalar masses obtained 
from different
ensembles 
are tabulated in
table~\ref{tab:mat}. In Fig.~\ref{fig:fpsLPS} we show the pseudoscalar 
decay constant computed from the valence overlap
fermions at the matching mass as a function of $a^2$. As in the case of 
free fermions, we indeed find a scaling of the pseudoscalar decay 
constant compatible with the expected $O(a^2)$ behaviour. 
However, we encounter a problem with this naive scaling test 
when extrapolating the mixed action pseudoscalar decay constants
to the continuum limit, we find $r_0 \fpsov=0.236(9)$ 
while the continuum extrapolated 
value for unitary twisted mass fermions comes out to be $r_0 \fpstm=0.181(10)$.

Thus, it seems that we are obtaining an inconsistent continuum limit
from our mixed action simulation. Of course, this observation demands
an explanation, since this is clearly in contradiction to our
expectation and must be an artifact of our setup. Indeed, the setup of
our naive scaling test is too naive. It does not take into account the
above mentioned mismatch of the eigenvalue spectra of the lattice
Dirac operators. In particular, the pseudoscalar correlator might be
strongly affected by zero modes of the overlap Dirac operator,
especially in the small volume and at the small value of the quark
mass employed by using ensembles $B_{\ell,V_1}$, $C_\ell$ and
$D_\ell$.

The suspicion of the overlap zero modes being responsible for the 
apparent inconsistent continuum limit values for the pseudoscalar 
decay constant leads to an {\em improved scaling test}.

\subsection{Improved Scaling Test}
\label{sec:improved}

In order to check the assumption that the zero modes play a special 
r\^ole in our mixed action setup, we should use observables for 
which such zero mode contributions are cancelled and which are therefore
in this sense ``improved''. 
Such observables can then be used both in the matching condition
and to compute the physical quantity. 
As we will demonstrate below, once an improved version of 
--in our case-- $\fps$ is used, all matching conditions 
considered lead to completely consistent continuum limit values. 

Our improved scaling test consists of 
studying instead of the pure 
pseudoscalar correlation function the correlator 
\begin{equation}
  \label{eq:PS}
  C_{\rm PP-SS}(t)\equiv C_{\rm PP}(t)-C_{\rm SS}(t)\,,
\end{equation}
where the scalar correlation function is subtracted and hence the
contribution of zero modes is exactly cancelled, since they are
eigenfunctions of $\gamma_5$.  In this way, we match the overlap
(mixed action) pseudoscalar meson mass resulting from such correlator
to a unitary pseudoscalar meson mass extracted from the standard PP
correlator.  This defines the {\em improved matching condition}.
Moreover, we extract $\fpsov$ from the PP-SS correlator.  Therefore,
we have an improved setup in both the matching condition and in the
observable itself -- in both we avoid the effects of zero modes.

\begin{figure}[t]
  \centering
  \subfigure[\label{fig:matLimpr}]{
    \includegraphics[width=0.33\linewidth,angle=270]{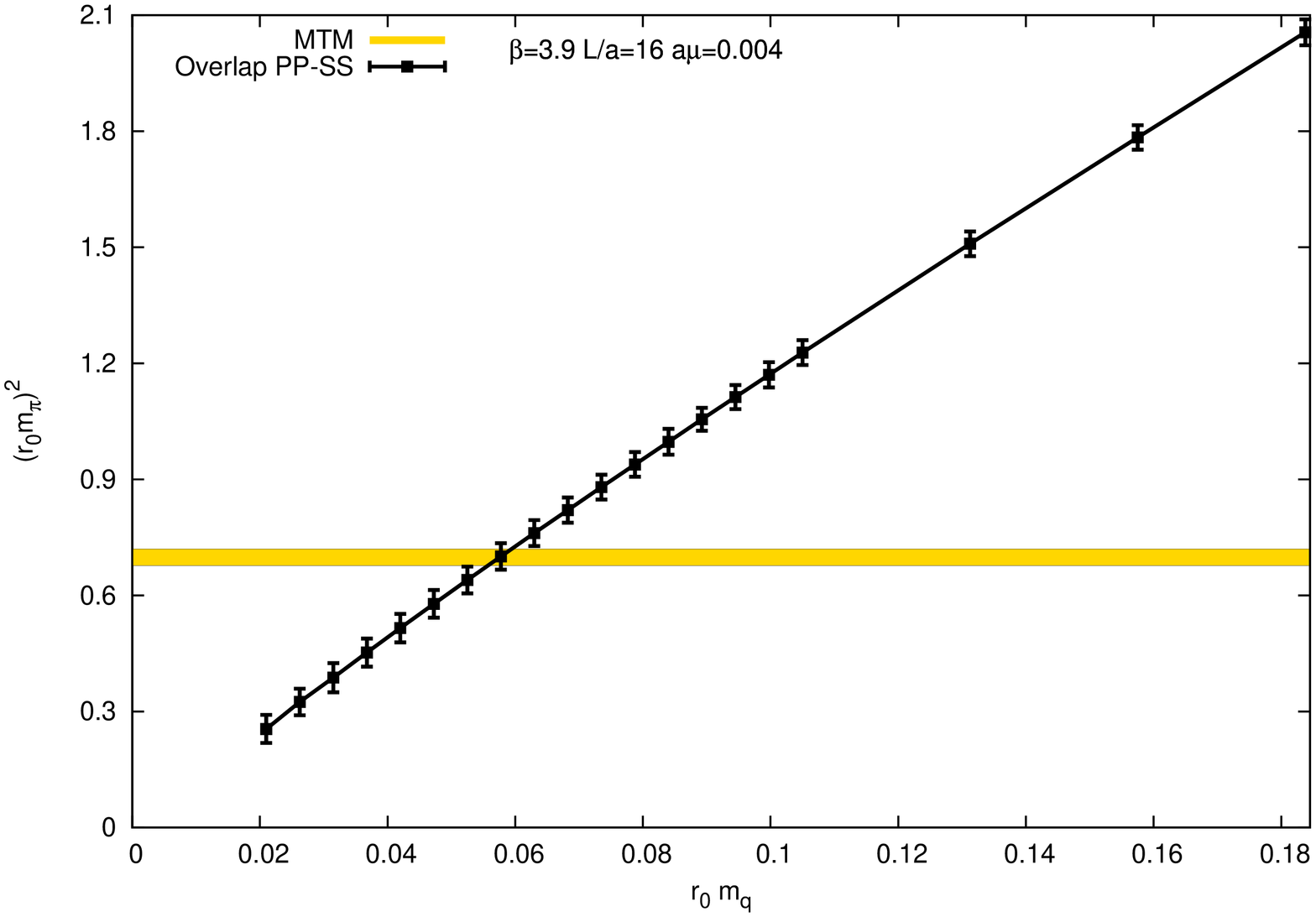}
  }
  \subfigure[\label{fig:fpsLPSimpr}]{
    \includegraphics[width=0.33\linewidth,angle=270]{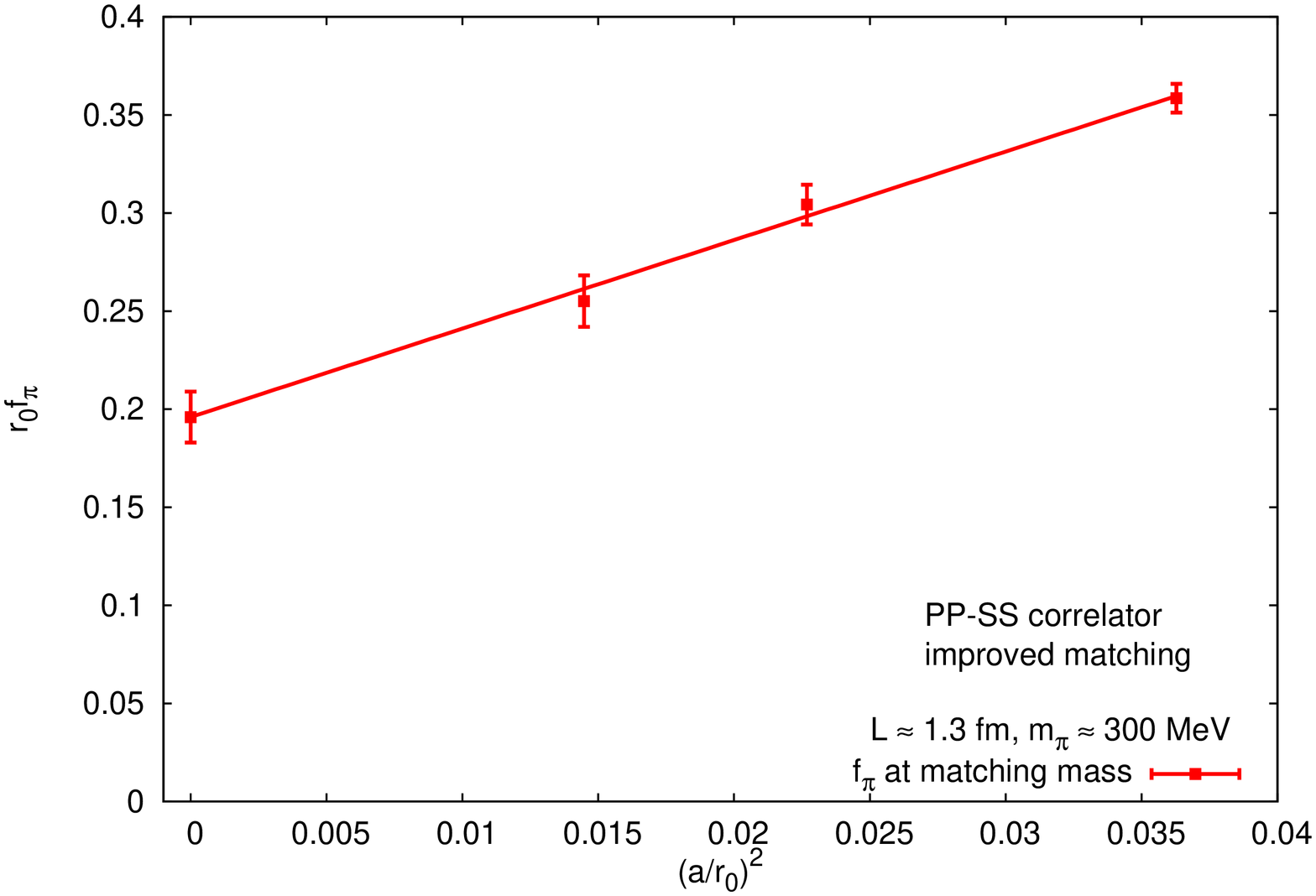}
  }
  \caption{(a) Improved matching condition for the case of ensemble $B_{\ell,V_1}$. (b) Continuum limit of $\fps$ in the case of the improved matching condition.  }
\end{figure}

We show the matching of the pseudoscalar mass 
in Fig.~\ref{fig:matLimpr} and the scaling in $a^2$ of the pseudoscalar 
decay constant in Fig.~\ref{fig:fpsLPSimpr} where we again observe the expected $\Oasq$ scaling behaviour. 
Performing a continuum limit extrapolation of $\fps$, we find $r_0 \fpsov=0.196(13)$ 
which is now completely compatible with the one of twisted mass 
fermions, $r_0 \fpstm=0.181(10)$.

\subsection{Alternative Scaling Test}

\begin{figure}[t]
  \centering
  \subfigure[\label{fig:matLalt}]{
    \includegraphics[width=0.33\linewidth,angle=270]{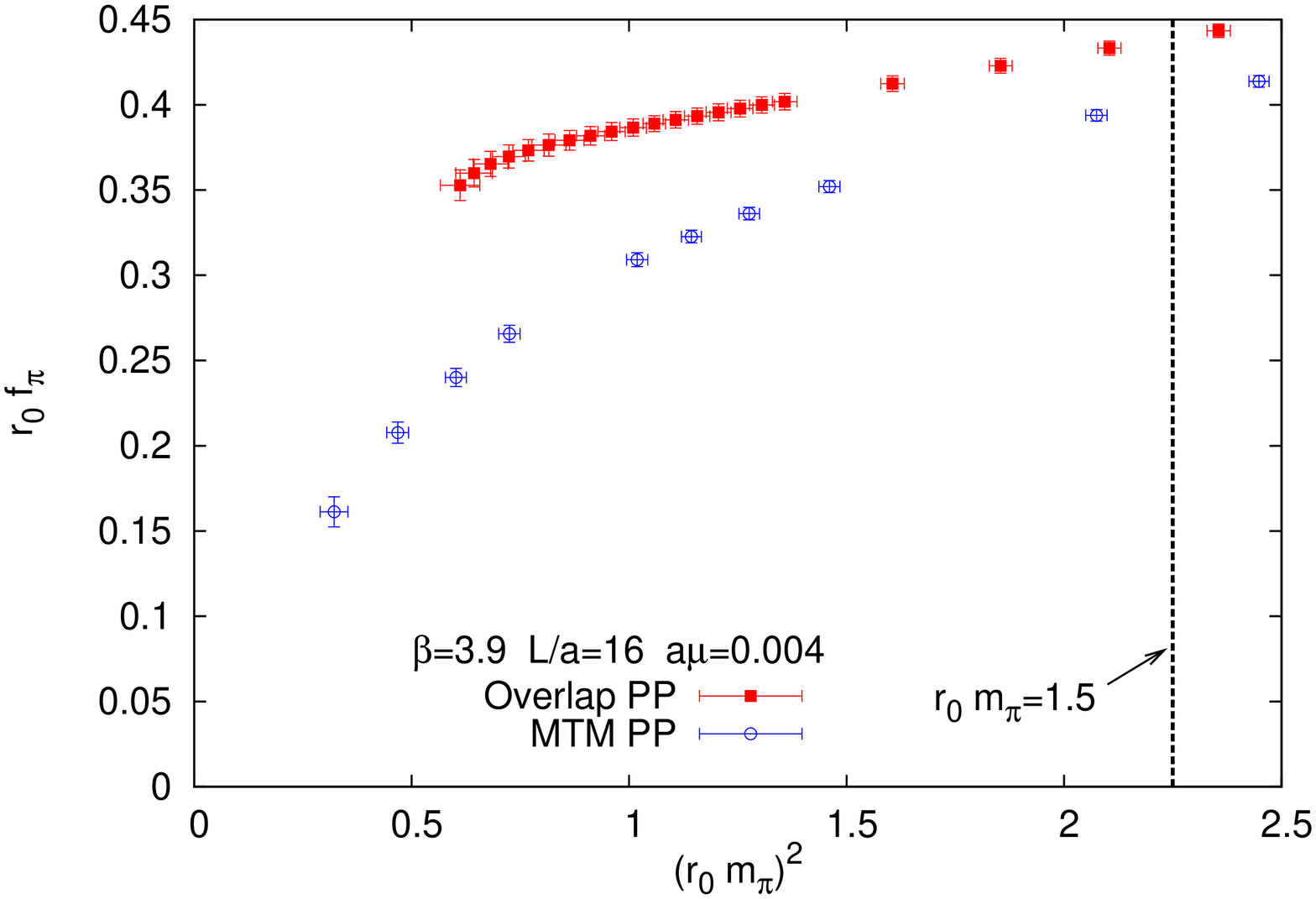}
  }
  \subfigure[\label{fig:fpsLPSalt}]{
    \includegraphics[width=0.33\linewidth,angle=270]{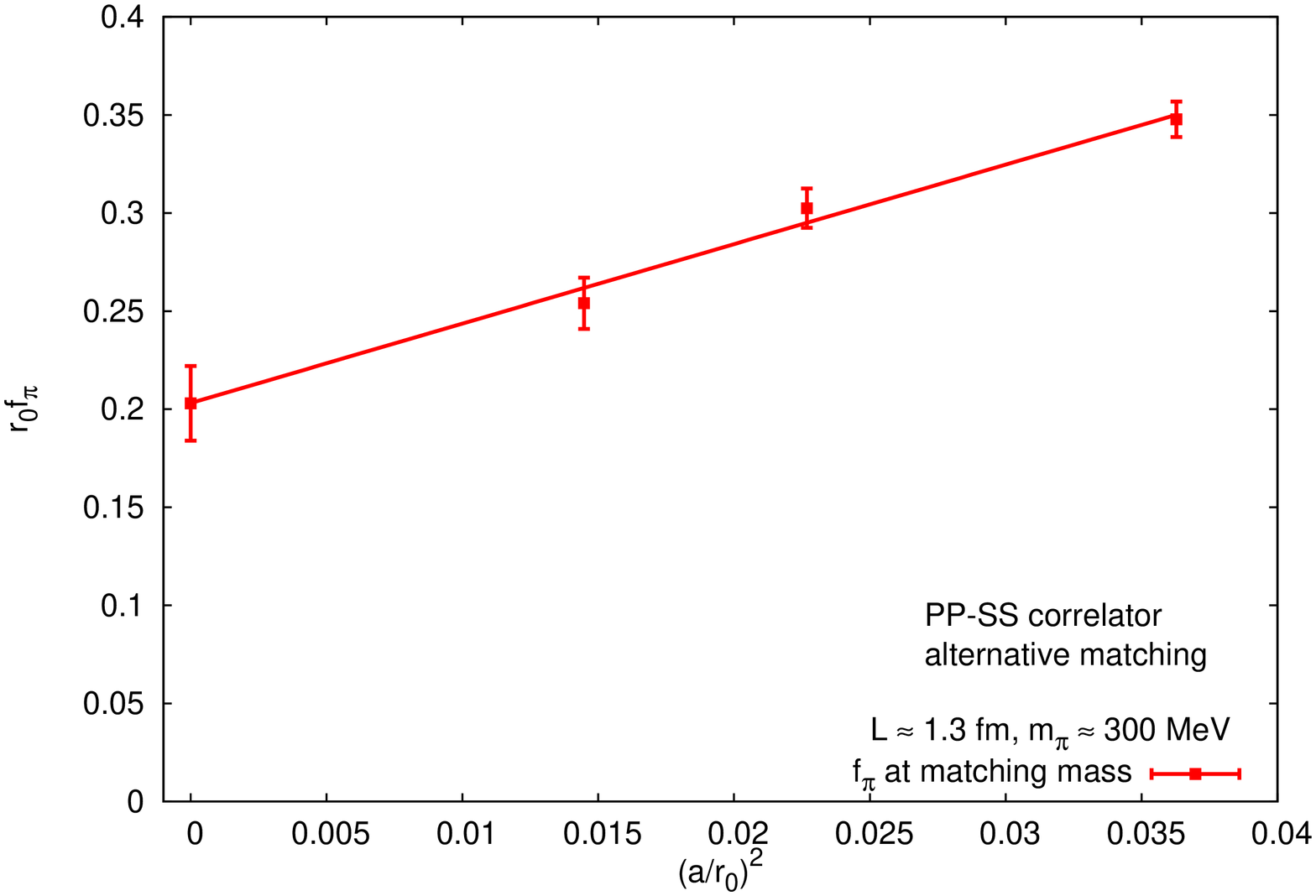}
  }
  \caption{(a) Matching of overlap and MTM quark masses (ensemble $B_{\ell,V_1}$) at a heavy quark mass value (yielding $r_0\mps=1.5$, which is denoted by vertical line), where the zero modes effects are strongly suppressed. Idea of the matching is explained in the text. (b) Continuum limit of $\fps$ in the case of the alternative matching condition.  }
\end{figure}

The results of the previous two sections indicate that indeed 
the zero modes play a special r\^ole in the mixed action setup 
used here. 
Using 
the correlator 
$C_{\rm PP-SS}(t)$ which cancels the zero mode contributions
exactly is not the only possibility to reach this goal. 
Our alternative scaling test starts by the observation that 
zero modes 
can also be suppressed by using either a large enough volume
or large quark masses, see also the discussion below. 
One way of avoiding the effects of chiral zero modes, at 
least in the matching condition, is then to
match the theories at a heavy quark mass values where the 
effect of the zero modes is strongly suppressed. 

An idea of such matching is the following.  We choose a value of $r_0
\mps=1.5$ in the regime of quark masses where the difference between
the $C_{\rm PP}(t)$ and $C_{\rm PP-SS}(t)$ correlators is negligible,
which indicates that the effects of the zero modes are small.  The
dependence of the pseudoscalar mass on the overlap and MTM quark mass
(using partially quenched MTM data) allows us to find the values of
quark masses which correspond to the chosen value of $r_0\mps$.  As
can be seen on Fig.~\ref{fig:matLalt}, the smallness of the zero modes
effects at this value of $r_0\mps$ means that the difference
$r_0(\fpsov-\fpstm)$ is significantly reduced with respect to the
large difference in $r_0\fps$ in the small quark mass regime.  The
ratio of MTM and overlap bare quark masses that lead to $r_0\mps=1.5$
provides an estimate of the ratio $R_{Z_P}\equiv Z_{\rm P}^{\rm
  ov}/Z_{\rm P}^{\rm tm}$.  In this way, we can define the {\em
  alternative} matching mass as $R_{Z_P}\mu_q$, where $\mu_q$ is the
unitary MTM quark mass.  The values that we find from this procedure
(for all ensembles with linear lattice extent $L\approx1.3$ fm) are
again tabulated in table~\ref{tab:mat}.

Once we have the matching masses from the alternative matching
condition, we can perform the continuum limit scaling of the
corresponding decay constant, extracted again from the $C_{\rm
  PP-SS}(t)$ correlator, such that the zero modes effects are
suppressed both in the matching condition and in the values of $\fps$
(in this sense, such setup is again an improved one with respect to
the naive scaling test).  Needless to say that we again observe a nice
$O(a^2)$ scaling of the pseudoscalar decay constant towards the
continuum limit (Fig.~\ref{fig:fpsLPSalt}), where we find $r_0 \fpsov
= 0.203(19)$, which is to be compared to the continuum limit value of
twisted mass fermions $r_0 \fpstm = 0.181(10)$. Again, a consistent
result is found. Note that a similar conclusion is obtained if instead
of performing the alternative matching with partially quenched data,
we use the unitary ensembles $B_h$, $C_h$ and $D_h$, at the heavier
masses, to impose the matching condition.

\section{Interpretation and Discussion}

\begin{figure}[t]
  \centering
    \includegraphics[width=0.33\linewidth,angle=270]{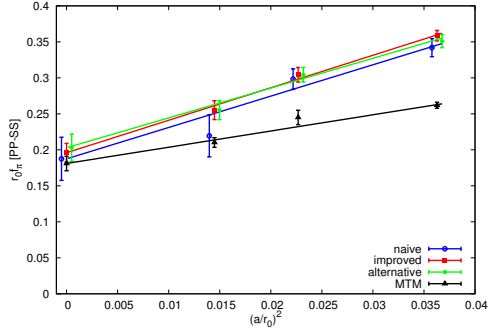}
  \caption{\label{fig:scalings} Continuum limit scaling of $r_0\fps$, using different matching 
conditions and extracting $\fps$ from the PP-SS correlator. All matching conditions 
lead to a consistent continuum limit with the one of the unitary approach. In the graph the data 
for ``naive matching'' is shifted to the left and the data for ``alternative matching'' 
to the right, for reasons of a better presentation.}
\end{figure}

In the previous section, we have discussed continuum limit scaling 
tests, employing three different matching conditions and using two 
definitions for $\fpsov$. 
The disturbing observation in these scaling tests has been that 
in case of the naive scaling test, an apparent inconsistency of the 
continuum limit value of $\fps$ has been encountered which, however, could be 
resolved by employing an improved and an alternative scaling strategy. 
One may wonder then, whether in the naive scaling test the matching condition, 
the used observable or even some interplay between both are responsible
for the inconsistent continuum limit of $\fps$.

Fig.~\ref{fig:scalings} provides an answer to this question, at least
in the case of $\fps$ considered here. In the graph, we use the naive,
improved and alternative matching conditions but evaluate $\fpsov$, in
all cases, from $C_{\rm PP-SS}(t)$. Obviously, with all three matching
conditions we obtain a completely consistent continuum limit value of
$\fpsov$, agreeing well with the one from the unitary setup. Thus, it
seems that the choice of the observable has a significant impact on
the observed mismatch in the naive scaling test of $\fps$ due to the
zero modes effects.

It is interesting to also perform the opposite test by using again
all three matching conditions but this time only use $\fpsov$ computed from 
the PP correlator. In this case, we obtain inconsistent 
continuum limit values, $r_0 \fpsov=0.236(9)$, $r_0 \fpsov=0.252(10)$ and $r_0 \fpsov=0.256(10)$ 
for the naive, improved and alternative matching conditions, 
respectively, while for the unitary setup we have the 
value $r_0 \fpstm = 0.181(10)$. 
As a consequence, 
it is important to define observables such that they are not 
influenced by zero modes. 

As we have shown above, mixed actions involving chiral fermions only in the valence
sector can exhibit particular effects. 
Indeed, the spectra of overlap and twisted mass Dirac
operators differ in an essential way. The overlap Dirac operator,
being chirally symmetric, incorporates the physical effects of exact
chiral zero modes at any value of the lattice spacing. This is not the
case for the Wilson twisted mass Dirac operator. In the context of a
mixed action this implies that the effect of zero modes in the valence
sector is not properly suppressed by a fermionic determinant which
depends only on sea quarks. Zero mode effects can therefore introduce
potential difficulties when comparing the continuum limit
extrapolation of mixed action data with respect to the unitary case.

The effects of zero modes are observable dependent. In particular, when
aiming at studies of QCD on a large volume, an appropriate choice
of the interpolating operator can be used to reduce or to cancel the
effects of zero modes on a given observable. Indeed, zero modes effects
can be interpreted as finite size effects and are thus suppressed as
the volume is increased. Moreover, when the quark mass increases, the
relative contribution of the zero modes decreases with respect to the one
of the other low-lying modes. Observables receiving contributions from
zero modes are therefore less influenced by their effects as
the quark mass increases.

Before closing this section, we collect a few remarks about the
continuum limit scaling of the mixed and unitary actions.

\begin{itemize}

\item The previous sections and discussion have illustrated the 
influence of exact zero modes of the overlap operator in a mixed action
setup. We have examined three examples of continuum limit scalings 
which give partially different results. Of course, our
list is by no means exhaustive, nor do we
claim to have provided the ``optimal'' continuum limit strategy. However, 
we have clearly   
provided a warning that in a mixed action setup special care 
has to be taken when lattice Dirac operators with exact chiral zero modes
are employed in the valence sector. In the following, we will make an attempt
to provide a quantitative estimate in the region of volume and 
pseudoscalar mass where the zero modes are the dominant effect and 
where they can be safely neglected. 

\item The $C_{\rm PP-SS}(t)$ correlator in eq.~(\ref{eq:PS}) has
  different discretization effects than $C_{\rm PP}(t)$. Indeed, it is
  known that the scalar correlator $C_{\rm SS}(t)$ is especially
  vulnerable to the double pole contribution to the meson
  propagator~\cite{Golterman:2005xa}. This unphysical effect is
  present in all unitarity violating setups (quenched, partially
  quenched and mixed action) and a separate analysis indicates that
  the scalar correlator indeed obtains a negative contribution of this
  kind, which has opposite sign with respect to the one from zero
  modes~\cite{Drach:2010}. 

\item The $C_{\rm PP-SS}(t)$ correlator receives excited state
  contributions from the scalar correlator. This can affect the
  extraction of the pseudoscalar ground state in the large quark mass
  region. In the regime of quark masses that we consider in this work
  this appears to be a negligible effect. In general, we observe clear
  signals for the plateaux, which we have furthermore controlled by
  comparing the extracted values of $\mps$ and $\fps$ to an analysis
  including an estimate of the systematic effect due to the choice of
  the fit interval, see also \cite{Drach:2010}.

\item For twisted mass fermions, the use of $C_{\rm PP-SS}(t)$,
  instead of $C_{\rm PP}(t)$, has a negligible impact on the values of
  $\fps$ and $\mps$ in the regime of parameters (quark mass, lattice
  size and lattice spacing) that we consider in this work. We only
  consider the latter case.

\item In the Wilson twisted mass formulation, a potential contribution
  from the large discretization effects present in the neutral pion
  mass can occur, in particular through finite volume
  effects. However, these cut-off effects are in principle properly
  taken into account by the fact that both the quark mass and the
  volume are kept fixed during the continuum limit extrapolation. The
  fact that we observe the expected $\Oasq$ scaling behaviour, gives
  us confidence that potential isospin breaking effects are indeed
  removed in the continuum limit, as also observed
  in~\cite{Baron:2009wt}.

\end{itemize}

\section{Quark Mass Dependence and Finite Size Effects}

In this section, we will perform a matching of the twisted mass sea 
and valence overlap actions at a heavier quark mass and at larger 
volumes. The purpose of this investigation is twofold. First of 
all, these investigations will tell us on a quantitative level 
at which value of the quark mass and volume the effects of the 
exact chiral zero modes are harmless (safe region), when their effect
is significant (hazardous region) and when their effect is unacceptably 
large (non-safe region). 
Second, using larger quark masses and volumes serves as a test of 
the picture developed in the previous sections about the r\^ole of the
zero modes in a mixed action simulation. 

\subsection{Quark Mass Dependence}

\begin{table}[t!]
  \centering
  \begin{tabular*}{1.0\textwidth}{@{\extracolsep{\fill}}lccc}
    \hline\hline
    Ensemble & $r_0\fps$ & $r_0\fps$ & $r_0\fps$\\
     & unitary MTM & naive matching & improved matching\\
     & PP correlator & PP correlator & PP-SS correlator\\
    \hline\hline
    $B_h$      & 0.3413(33) & 0.3973(65) & 0.3626(107)\\ 
    $C_h$      & 0.3373(33) & 0.3774(63) & 0.3669(78) \\
    $D_h$      & 0.3382(50) & 0.3493(100)  & 0.3558(90) \\
    \hline
    continuum limit & 0.334(11) & 0.330(17) & 0.345(18)\\
    \hline\hline\\
  \end{tabular*}
  \caption{The values of the pseudoscalar decay constant for ensembles $B_h$, $C_h$
    and $D_h$ at the matching mass (with the naive and improved
    matching condition). The continuum limit values are also shown.}
  \label{tab:heavy}
\end{table}

For the matching of the heavier quark mass we use the ensembles $B_h$,
$C_h$ and $D_h$ of table \ref{tab:setup}.  We apply for these
ensembles both the naive matching procedure (naive matching condition
using $\fps$ from the PP correlator) and the improved one (improved
matching condition using $\fps$ from the PP-SS correlator).  In both
cases we observe, as expected, a nice scaling of the pseudoscalar
decay constant in $a^2$ towards the continuum limit. In table
\ref{tab:heavy} we give the values of the pseudoscalar decay constants
from the naive and the improved matching conditions, as well as one
for the unitary twisted mass case. As can be seen in the table, the
differences in the values of $\fpsov$ and $\fpstm$ for all ensembles
are much smaller than in the case of $B_{l,V_1}$, $C_l$ and $D_l$
ensembles and all continuum limit values are consistent with one
another.  This clearly confirms our interpretation that the zero modes
effects are large at a quark mass corresponding to a pseudoscalar
mass of $300$ MeV (in infinite volume).  At $450$ MeV, these effects
are strongly suppressed although not totally negligible (see table
\ref{tab:heavy}).

\subsection{Finite Size Effects}

Finite size effects (FSE) in the light pseudoscalar meson observables
considered in this work can have multiple origins. As mentioned
earlier, zero modes of the Dirac operator can introduce significant
finite volume effects in the pseudoscalar correlator. Other sources of
FSE are related to pion loops or to the actual size of the box with
respect to the typical hadronic scale. Concerning pion loops, in the
case of twisted mass fermions a contribution to FSE from the lighter
neutral pion mass can in principle be present at finite lattice
spacing, although this is expected to vanish when performing the
continuum limit extrapolation (as discussed in the previous
section). When considering the scalar correlator, an additional FSE
arises from the bubble diagram describing the exchange of two
pseudoscalar mesons. Isolating the relative contribution of these
effects in our data for the mixed and unitary actions is beyond the
scope of this work. Here, we rather concentrate on the comparison of
the dependence of $\fpstm$ and $\fpsov$ on the volume, since this
should give some evidences of the presence of FSE from the
topological zero modes.

We consider three ensembles $B_{\ell,V_1}$, $B_{\ell,V_2}$,
$B_{\ell,V_3}$ at the coarsest value of the lattice spacing, see
table~\ref{tab:setup}. The corresponding linear lattice extents are
$1.3\,\fm$, $1.7\,\fm$ and $2.0\,\fm$, respectively. The quark mass is
set to $m_{q} \approx 20\,\mev$ and the values of $\mps L$ at the
matching point are $2.5$, $2.8$, $3.3$, for ensembles $B_{\ell,V_1}$,
$B_{\ell,V_2}$, $B_{\ell,V_3}$, respectively.

\begin{figure}[t]
  \begin{center}
  \subfigure[\label{fig-fse-fps}]{
    \includegraphics[width=0.33\textwidth,angle=270]{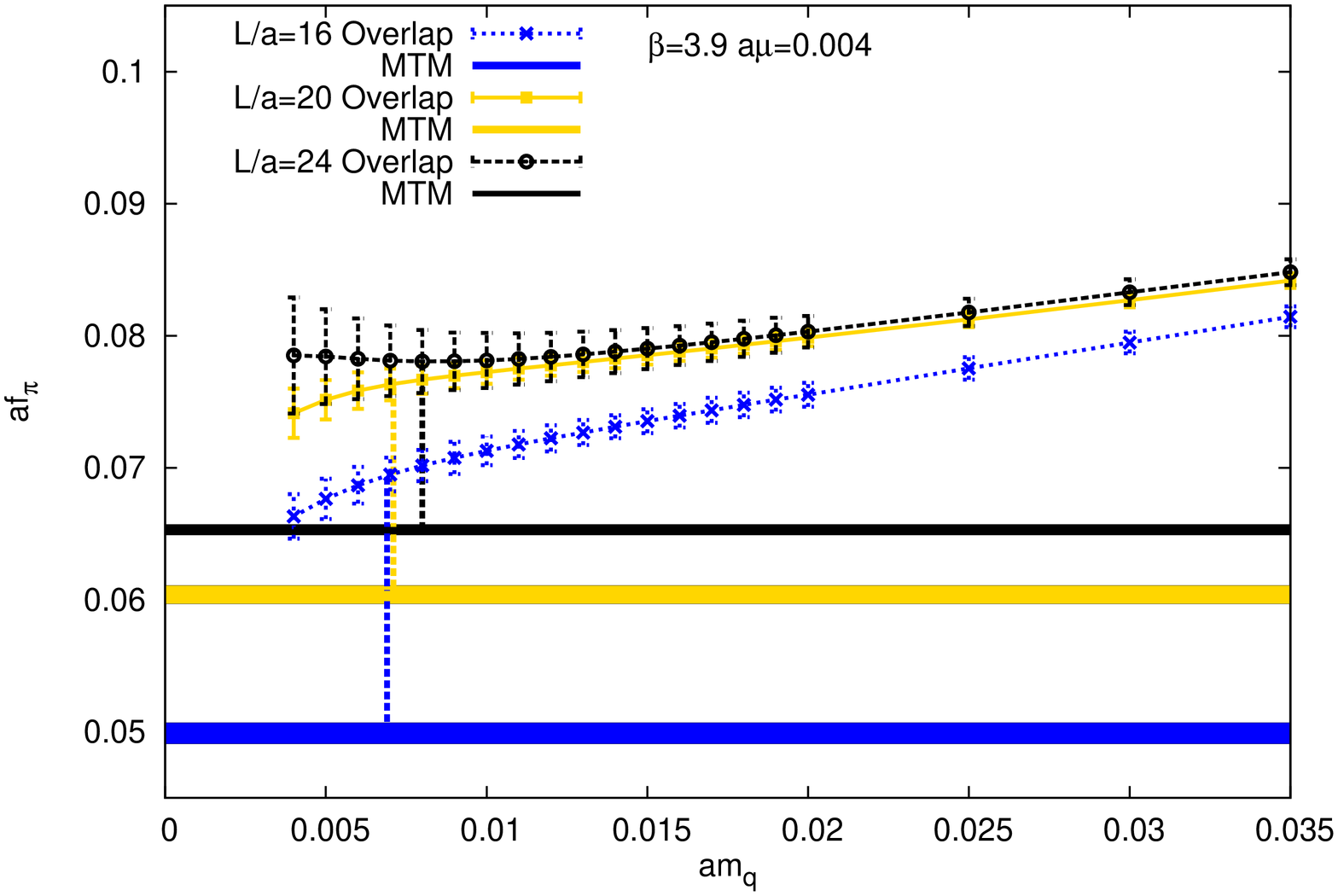}
  }
  \subfigure[\label{fig-fse-fps-diff}]{
    \includegraphics[width=0.33\textwidth,angle=270]{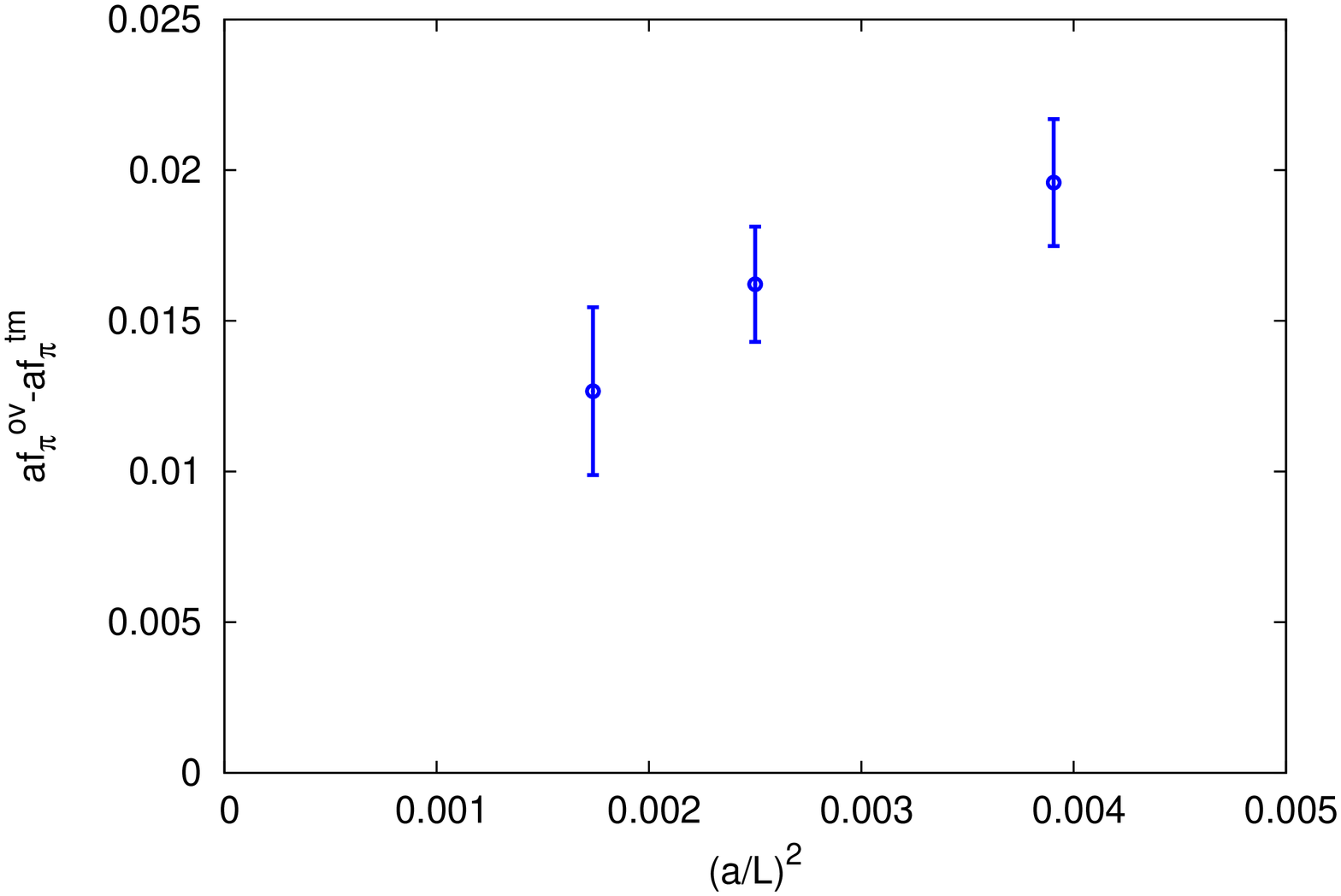}
  }
  \end{center}
  \caption{Finite size effects from ensembles $B_{\ell,V_1}$,
    $B_{\ell,V_2}$, $B_{\ell,V_3}$ (see table~\ref{tab:setup}). (a)
    Quark mass dependence of the pseudoscalar meson decay
    constant. (b) Volume dependence of the difference between the
    pseudoscalar meson decay constant in the mixed and unitary
    actions. We observe that the difference between $\fpstm$ and
    $\fpsov$ decreases when increasing the volume. }
  \end{figure}

To illustrate the magnitude of finite size effects, we plot the quark
mass dependence of the pseudoscalar meson decay constant -- in this
section we only use the PP correlator -- for our three ensembles in
Fig. \ref{fig-fse-fps}. The horizontal bands correspond to the unitary
values of $\fpstm$. The matching mass (employing the naive matching
procedure) is shown by the vertical dashed line. We observe
significant FSE for both the mixed and unitary actions when reducing
the lattice size from $1.7\,\fm$ down to $1.3\,\fm$, while they are
smaller between $2.0\,\fm$ and $1.7\,\fm$. Note that for $\fps$ there
is a relative sign in the contribution to FSE coming from pion loops
and zero modes. This could induce, in practise, some cancellations of
FSE in the mixed action data.

The differences between the decay constants $\fpsov$ and
$\fpstm$, at the matching mass, are plotted in
Fig. \ref{fig-fse-fps-diff}. In such difference, many of the
contributions to FSE are expected to cancel. The fact that
$\fpsov-\fpstm$ has a non vanishing slope when changing the lattice
size can be interpreted as a FSE coming from the zero modes of the
overlap operator. This hypothesis is supported by the fact that this
slope is within errors compatible with zero when instead of using
$\fpsov[{\rm PP}]$ in the difference, one considers $\fpsov[{\rm
    PP-SS}]$, extracted from the $C_{\rm PP-SS}(t)$ correlator.

Fig.~\ref{fig-fse-fps-diff} provides indications that FSE from
topological zero modes are not totally negligible for the ensemble
$B_{\ell,V_3}$, with our largest lattice size $L\approx 2.0\,\fm$ and $\mps L
\approx 3.3$. For the heavier ensemble $B_h$, with $L \approx
1.3\,\fm$ and $\mps L \approx 3.1$ and the same value of the lattice
spacing, we observed that $a\fpsov-a\fpstm=0.011(1)$, arising
mainly from unitarity violations, since zero mode effects were observed
to be small in this case. By taking this value as a rough estimate for
the size of unitarity violations at this value of the lattice spacing,
we estimate that a lattice size $L \geq 2.4\,\fm$ (corresponding to
$\mps L \geq 4$) would be needed, at the lighter quark mass
$m_{q}\approx 20\,\mev$, to enter in the regime of small FSE from
chiral zero modes. In practise, this would require to deal with the
overlap operator on an $L/a=32$ lattice.

We close this section by a remark about the size of the unitarity
violations in our mixed action. The difference $\fpsov-\fpstm$ in
Fig. \ref{fig-fse-fps-diff} is numerically large for all three
volumes. This difference is not expected to vanish in the infinite
volume limit due to the presence of unitarity violations in the mixed
action data at non-zero values of the lattice spacing. At a value of
the lattice spacing, $a \approx 0.08\,\fm$, with $L\approx 2.0\,\fm$
and $\mps L \approx 3.3$, we observe that the mixed and unitary values
of $\fps$ can differ by $20\%$. Fig.~\ref{fig:scalings} illustrates a
similar effect on a smaller volume. Indeed a significant difference
between overlap and twisted-mass data is observed at the coarser value
of the lattice spacing, $a \approx 0.08\,\fm$. This points towards the
necessity of a proper control of the continuum limit extrapolation
when working in a mixed action setup.

\section{Conclusion}

We have studied the continuum limit scaling of a mixed action with
overlap valence quarks on a twisted mass sea employing three values of
the lattice spacing. We observe the expected $O(a^2)$ scaling
behaviour for the here investigated pseudoscalar decay constant.  We
have demonstrated that in such mixed action approach the zero modes of
the overlap operator which are unmatched by the twisted mass operator
at the lattice spacings employed here, can play a special r\^ole. They
can severely affect physical quantities such as the pseudoscalar decay
constant.

\begin{figure}[t!]
  \begin{center}
    \includegraphics[width=0.45\textwidth,angle=270]{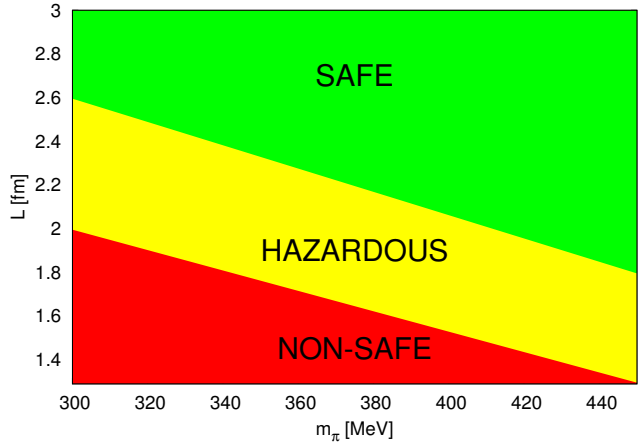}
    \caption{A qualitative graph indicating the safe,
      hazardous and non-safe regions of parameters in physical units
      (linear extent of the lattice vs. pseudoscalar meson mass) in
      mixed action simulations with overlap valence and twisted mass
      sea quarks. The parameter values in the ``safe'' region are such
      that the effects of chiral zero modes of the overlap operator
      are negligible. The boundary between the ``safe'' and
      ``hazardous'' region corresponds to $\mps L=4$ and the
      boundary between ``hazardous'' and ``non-safe'' is $\mps L=3$.}
    \label{fig-safe}
  \end{center}
\end{figure}

The precise values of the quark mass and/or lattice size at which the
effects of zero modes become relevant for a given observable are not
known a priori. A dedicated numerical study is therefore needed to
address this question.  To this end, we performed simulations at two
values of the quark mass and three volumes.  These calculations then
provided information on the values of the parameters for which these
zero mode effects are significant or when they can be neglected.  Our
findings are summarised in Fig.~\ref{fig-safe}. There we indicate 
on a more qualitative level the
regions in the quark mass and volume where simulations ought to be
safe, i.e. where the zero modes play no significant r\^ole, the
hazardous regions, where their effect can be significant and a
non-safe region, where their effect become really large.

We want to stress that the difficulties in the continuum limit
approach of $\fps$ arising from 
the zero modes of the overlap Dirac operator originate solely 
from the particular definition of $\fps$ and not from the 
matching condition employed. We demonstrated this clearly by, 
employing three different matching conditions, but using 
only a definition of $\fpsov$ (constructed from 
$C_{\rm PP-SS}(t)$) which is not affected by zero modes, 
see Fig.~\ref{fig:scalings}. 
On the contrary, 
by employing a definition of 
$\fpsov$ which is highly sensitive to zero modes
(i.e. constructed from $C_{\rm PP}(t)$), we find 
inconsistent continuum limit results for all three 
matching conditions employed. 
This might suggest that if it is possible to construct physical 
observables such that they are not affected by zero modes
consistent continuum limit values can be obtained.
However, this expectation needs further corroboration by studying
other physical quantities.
 
The effects of zero modes depend on the choice of the
operator. Furthermore, their effects vary from one observable to
another. We refer to a forthcoming publication~\cite{Drach:2010} for a
study of the impact of zero modes on other observables and
operators. For some observables, isolating the contribution from zero
modes can be a difficult task. For the study of such observables, it
is particularly important to perform the simulations in a regime of
parameters where these effects are negligible.

By working in the safe regime, where the effects of 
zero modes are negligible, we believe that this mixed
action approach should allow to obtain accurate determinations of
quantities for which good chiral properties of valence fermions are
essential, such as the kaon bag parameter $B_K$ or the amplitudes of
$K\rightarrow{\pi}{\pi}$ decays.

\subsubsection*{Acknowledgements}

We gratefully acknowledge valuable discussions with Vincent Drach,
Elena Garc\'ia Ramos, Maarten Golterman, Carlos Pena, Luigi Scorzato and
Piotr Tomczak. We thank the ETM collaboration for providing us with
the dynamical gauge field configurations used in this work. This work was partially
supported by the DFG Sonderforschungsbereich / Transregio SFB/TR-9.
The computer time for this project was made available to us by the Leibniz
Rechenzentrum in Munich on the HLRB II system. We thank this computer
centre and its staff for all technical advice and help. We also
acknowledge the use of computer resources of the Poznan Supercomputing
and Networking Centre (PCSS).  K.C. was supported by Ministry of
Science and Higher Education grant nr. N N202 237437.

\end{document}